\documentclass[a4paper,fleqn,usenatbib]{mnras}

\usepackage[T1]{fontenc}
\usepackage{ae,aecompl}

\usepackage{graphicx}	
\usepackage{amsmath}	
\usepackage{amssymb}	
\usepackage{natbib}

\title[BEANS -- a
software package for distributed Big Data analysis]{BEANS -- a
software package for distributed Big Data analysis}

\author[Arkadiusz Hypki]{
Arkadiusz Hypki,$^{1,2}$\thanks{E-mail: ahypki@strw.leidenuniv.nl}
\\
$^{1}$Leiden Observatory, Leiden University, PO Box 9513, NL-2300 RA
Leiden, the Netherlands\\
$^{2}$Nicolaus Copernicus Astronomical Center, Bartycka 18, 00--716 Warsaw,
Poland
}

\date{Accepted XXX. Received YYY; in original form ZZZ}

\pubyear{2016}

\begin{document}
\label{firstpage}
\pagerange{\pageref{firstpage}--\pageref{lastpage}}
\maketitle

\begin{abstract}

BEANS software is a web based, easy to
install and maintain, new tool to store and analyse data in a distributed way for a massive amount of
data. It provides a clear interface for querying, filtering, aggregating, and
plotting data from an arbitrary number of datasets. Its main purpose
is to simplify the process of storing, examining and finding new
relations in the so-called Big Data.

Creation of BEANS software is an answer to the growing needs of the astronomical community 
to have a versatile tool to store, analyse and compare the complex astrophysical
numerical simulations with observations (e.g. simulations of the Galaxy or
star clusters with the Gaia archive).
However, this software was built in a general form and it is ready to use in any other
research field or open source software. 

\end{abstract}




\section{Introduction}
\label{sec:Intro}

BEANS\footnote{\url{www.beanscode.net}} is an attempt to provide to the
community an integrated software which works as a central repository for
arbitrary amount of data and simultaneously as a platform with proper tools to
analyse the data and extract usable knowledge from it. BEANS can store, manage,
filter and aggregate any amount of data using tools already well verified and
proven in the industry. BEANS was originally designed to help astrophysicists to
analyse data from complex numerical simulations and compare them with
observations. However, BEANS is a general tool and it can be used with great
benefits by any users who deal with a large amount of data and who wish to have
an easy to understand platform for complex data analysis. BEANS has web and
console interfaces. The web interface allows to use BEANS from any device
equipped with a browser, whereas the command line interface ensures that the
data can be accessed also by the external software, if needed. BEANS is under a
heavy development. New features are being added continuously which should make
it even more useful for a wider range of users.

The amount of data in science which is delivered nowadays increases like never
before. It applies for all fields of research. However, physics and astrophysics
appear to push the requirements for data storage and analysis to the boundaries.
With already existing missions like Gaia \citep{Gaia2008IAUS..248..217L}, and
for future projects like LSST \citep{2008arXiv0805.2366I}, the need for reliable
and scalable data storage and management is even higher.

In order to manage data, and more importantly, to gain new knowledge from it, it
is crucial to have a proper toolbox. The amount of data for many projects is too
big for simple processing with bash or python scripts. Tools which would allow
to simplify data analysis are crucial in research these days. Equally important
is also to have a software to store the data. Flat files stored on a server
may suffice, but in a case of thousands of simulations which differ in initial
conditions, searching for a specific simulation may be a problem. Thus, it is
very important not only to store the data but also describe it with an
additional parameter (meta-data) in order to find it easily later.  

BEANS software was initially created for easy data management and analysis
of thousands of numerical simulations done with the MOCCA
code\footnote{\url{http://www.moccacode.net/}}. MOCCA is able to
perform numerical simulations of real-size globular star clusters and at the
same time it allows to track a full dynamical and stellar evolution of every
object in the system. MOCCA follows with great details the formation, dynamical
and stellar evolution of exotic objects like blue stragglers stars \citep{Hypki2013MNRAS.429.1221H},  
formation of intermediate mass black holes in GCs
\citep{Giersz2015MNRAS.454.3150G}, cataclysmic variables, compact
binaries, and more. MOCCA follows the star cluster
evolution closely to N-body codes \citep{Wang2016MNRAS.tmp...74W} but is
incomparably faster. In the same amount of CPU time one can have
many MOCCA simulations for various initial conditions and perform detailed
statistics of globular star clusters. Thus, it is so important to have a tool to
efficiently store, search and analyse a large number of simulations.

One MOCCA simulation on average provides around ten output files which take in
total around 10-20~gigabytes~(GBs), depending on the initial conditions of a
star cluster.
Simulations for a set of various initial conditions can easily exceed hundreds
of simulations and the requirement for data storage increases to
terabytes~(TBs).
Although it is much less than the needs of huge astrophysical surveys today, it
is still a lot of data which demands a proper set of tools to easy manage, query
and visualize data. BEANS is suitable for both, a large number of datasets as
well as large datasets. BEANS uses internally only widely used and well adopted
solutions from the industry. These technologies
(see~Sec.~\ref{sec:Technologies}) can handle petabytes~(PBs) of data stored on
hundreds or thousands of nodes. Thus, BEANS is suitable to analyse various
astrophysical problems including even the largest astronomical missions (e.g.
Gaia, LSST) and the largest numerical simulations.

Although the BEANS software was created to manage astrophysical simulations, it
is written in a very general form. It can be used in any field of research, or
another open source projects (see licence in Sec.~\ref{sec:License}). The only
requirements are: first, to have data in a tabular form, and second, to use the
Pig Latin language (see Sec.~\ref{sec:Technologies}).

This paper is organized as follows. In the Sec.~\ref{sec:Technologies} there is
a description of the technologies used by BEANS. The power of BEANS is at least
as good as the sum of the features of the underlying technologies. In the
Sec.~\ref{sec:BasicTerms} there are introduced the basic terms used in BEANS
which are necessary to understand the examples in the next
Sec.~\ref{sec:Examples}. There are three examples with increasing level of
complexity which show step by step how to use the BEANS software and how
powerful the data analysis with that software is. The next
Sec.~\ref{sec:Plugins} gives a general description on how to extend basic BEANS
functionalities with plugins. Sec.~\ref{sec:FuturePlans} presents the
overall development roadmap and the next main features scheduled for
implementation. The next Sec.~\ref{sec:License} specifies the selected
licence and the last Sec.~\ref{sec:Summary} summarizes the features of the BEANS
software.

\section{Underlying technologies}
\label{sec:Technologies}

In this section the main technologies which are used in the BEANS software are
presented. Their value was already well proven in the industry (e.g. by Google,
Yahoo). However, beyond
any doubts, they are equally powerful for science too.

BEANS software works in a server -- thin clients mode.
It means that all calculations are performed on a server, or more preferably,
clusters of servers.
The thin client is either a browser or a command line interface which defines
queries, plots, imports or exports of the data. However, the whole analysis is
performed on the server side. In this way a client can be any device equipped
with a web browser (desktop PC, laptop, tablet, phone, etc.). In order to
increase the computational power of the entire cluster, only the server side has
to be equipped with more machines. Moreover, updating BEANS to a newer version
can be applied on-the-fly only on the server side on the working cluster -- the
clients do not change.

The database which is used internally by the BEANS software is Apache
Cassandra\footnote{\url{http://cassandra.apache.org/}}. It is decentralized,
automatically replicated, linearly scalable, fault-tolerant with tunable
consistency database written in Java. As compared to well known relational
databases like MySql, it is designed from scratch for distributed
environments. Thus, Apache Cassandra (shortly Cassandra) solves a lot of
problems which relational databases were struggling with (e.g. data replication,
failovers of nodes). Cassandra itself can easily handle PBs of data scattered
across hundreds or thousands of machines. What is more important, one person
can easily manage medium size cluster of nodes with Cassandra.

Cassandra has a lot of features which are
extremely important, especially for users who might not have deep technical
knowledge. We discuss them in the following paragraphs.

Cassandra is easy to maintain and extend. Every person who has a moderate
understanding of Unix platform (e.g. Linux, Max, BSD)
 should have no problem with building and
maintaining a Cassandra cluster. Such a cluster can be expanded if there is a
need to have more space in a database. Any additions of new nodes to an already
existing Cassandra cluster is not more complicated than downloading the
software, changing a few things in the configuration file, and starting
Cassandra on a new node.
The Cassandra cluster will determine automatically that the new node arrived and
the cluster will eventually distribute the data in background equally over the
entire cluster. Thus, it is not needed for the user (e.g. astrophysicist) to
have a deep technical knowledge to take the full advantage of using BEANS
software.

Cassandra by default stores each piece of data on multiple machines (by default
three different copies on three different nodes). It means that even if some of
the nodes fail because of the hardware problems, the data are safe. One may
bring back the node back on-line without hurry and that action will not even
disturb the normal operation of the whole Cassandra cluster.

Cassandra scales linearly with the number of nodes. It means that one may
increase the overall computing power of the cluster just by increasing the
number of nodes. The queries against the data will run faster if the number
of nodes increases because the scripts will read the data from the closest copy
of the data.

Elastic\footnote{\url{https://www.elastic.co/}} (formerly Elasticsearch) is
the second major component of the BEANS software. It is a powerful open source search engine. It
is used by BEANS to provide full search capabilities for datasets, tables,
notebooks, and users. 
With Elastic it is easy for BEANS to instantaneously search for any interesting datasets or
tables (see~Sec.~\ref{sec:BasicTerms}).

Elastic and Apache Cassandra are examples of emerging NoSQL technologies.
NoSQL meant initially the databases which did not demand full details of data
schemas, like relational databases do. Majority of NoSQL databases were design
from scratch to work in the distributed environments. After some time NoSQL
solutions started to actually use, initially abandoned, SQL language (or
something similar) for querying. And the SQL databases started to move towards
the distributed solutions too. 

For BEANS the flexibility of NoSQL solutions are crucial. The data model can be
simpler and can be much easier altered in the future. Moreover, instead of
defining numerous different tables with different columns, one can have only one
table which can store any type of data. This simplifies development and allows
the BEANS software to store arbitrary number of different tables without
enforcing the database to handle internally thousands of separate tables.

One of the greatest advantages of using Apache Cassandra is its integration with
MapReduce paradigm \citep{Dean1327452.1327492}. MapReduce was designed to help
process any data in distributed manner and it consists of two stages. In the
first stage, called \textit{map}, data is read and transformed into new pieces
of information. In the second stage, called \textit{reduce}, data produced in
the first stage can be aggregated (with functions like sum, average, count,
etc.). 
This approach is perfect for problems which are easy to parallelize, it allows
to have linear scalability and works on commodity hardware. This last feature is
very important because it allows to create powerful computer clusters relatively
cheap. For details on MapReduce see \citet{Dean1327452.1327492} or Apache Hadoop
\footnote{\url{http://hadoop.apache.org/}} project which is an open-source
software for reliable, scalable, distributed computing based on the MapReduce
approach.

The MapReduce approach is a low level solution. Details on how MapReduce jobs
are processing the data are not important for the end-users. In order to use the
power of distributed data analysis more efficiently, there are available
higher-level languages and libraries. These libraries usually transform queries
internally to a sequence of MapReduce jobs. Among the most popular there are
Apache Pig, Hive, Cascading, and recently Apache Spark.
They represent different approach for the same problem. For example,
Hive\footnote{\url{http://hive.apache.org/}} is very similar in syntax to SQL
language. However, for the BEANS software, we decided to choose Apache
Pig\footnote{\url{http://pig.apache.org/}} project. Scripting language for
Apache Pig is called Pig Latin. The main reason for this choice is that in my
opinion scripts written in Apache Pig are much more intuitive and clear than in
any other high level MapReduce language. The script in Pig Latin consists of
simple series of instructions and even the larger scripts are still quite easy
to understand -- especially for users who did not write the script in the first
place. Whereas SQL-based languages (e.g. Hive) are declarative languages and the
complex scripts are considered as less intuitive and less readable. Recently,
the Spark project gained a lot of attention as being much faster for some
applications than pure MapReduce jobs and thus performs faster than Apache Pig
and Hive. There is already taken an effort by the developers of Apache Pig to
run computations actually as Spark jobs. In this way it will be possible to have
a clear scripting language like Pig Latin but working much faster using Spark
platform. Another solution to speed up computations is to use Apache
Ignite\footnote{\url{https://ignite.apache.org/}}. The alternatives will be
carefully evaluated and the optimisation to BEANS will be applied in the next
releases.

Pig Latin language should be easy to learn for any user with some programming
experience. The basic description of the Pig Latin, which are enough to start,
fit to one web-page\footnote{Pig Latin Basics for Apache Pig 0.15.0
\url{http://pig.apache.org/docs/r0.15.0/basic.html}}. The main advantage of
using Pig Latin is that it is a very simple scripting language which consists of
well known operations like \textit{C = A operator B;} -- data stored in
so-called aliases \textit{A} and \textit{B} are transformed with the
\textit{operator} and the result is saved to alias \textit{C}. It is very much
the same as for conventional languages (e.g. C, python, Java). The only
difference is that aliases \textit{A}, \textit{B} and \textit{C} represents
actually the data, not numbers or strings of characters. The basic concepts
behind the Pig Latin language is to transform the data from one form
(\textit{A}, \textit{B}) into another (\textit{C}).

Typical example of a statement in Pig Latin language is
grouping. Here the statement \textit{C = GROUP A BY age;}
allows to split all the rows under alias \textit{A} into a set of groups by the
column \textit{age}. The resulting grouped data are stored under the alias
\textit{C}. From that point the rows in alias \textit{C} one can transform into
another form, e.g. one can aggregate the rows in \textit{C} by
computing minimum, maximum or average values for every group separately;
one can find distinct values for every group; filter the rows for each groups;
join the rows with some other data, and many more. The bottom line is that Pig Latin
language defines series of statements which take the data as arguments and
transform it into a new form. It keeps the script easy to understand. 

Pig jobs can be executed in local and distributed mode. In the local mode all
the computations are performed on the computer which started the Pig job. This
might be enough for a small amount of data. However, to take the full
advantage of distributed computing one has to start Pig jobs on the Apache Hadoop
cluster. Hadoop is and open-source platform for reliable, scalable and
distributed computing. The overall computing power of Hadoop can be easily
extended by starting more Hadoop nodes. This would speed up the times for Pig
scripts to accomplish too.

An integral part of any software for data analysis is visualisation. In the
first public version BEANS provides only basic
plots like point or line plots and histograms.
Additionally, BEANS is being integrated with a number of plots produced with D3
library\footnote{\url{http://d3js.org/}}. The D3 library is an example
on how computer sciences and art sciences can work together creating
non-standard plots which at once are powerful, interactive, clean and easy to
read. Examples of D3 plots are presented in Sec.~\ref{sec:Examples}. New
types of plots and D3 visualisations will be added continuously with the next BEANS releases.

BEANS uses the full potential of the Representational state transfer (REST)
approach\footnote{\url{https://en.wikipedia.org/wiki/Representational_state_transfer}}.
There are many advantages of using REST approach for building complex software.
However, there are a few reasons which are especially important for software
like BEANS. 

The REST services work in a client-server mode. The client is very lightweight
and the server is very powerful. BEANS is designed to handle analysis of 
huge datasets. The clients are designed only to exchange
small portions of data (e.g. submit query, view data) using the REST interface.
Whereas all the heavy analysis is performed on the server side. If there is a need
to extend the overall computing power of the platform one can add more servers and
the clients would not even notice the difference -- the underlying server
infrastructure is hidden from the clients.

The REST services communicate using protocols which
are independent of the actual implementation. Usually the data are transferred
using JSON\footnote{\url{https://en.wikipedia.org/wiki/JSON}} or
XML\footnote{\url{https://en.wikipedia.org/wiki/XML}} format -- there are
completely independent of the database used on the server (e.g. Apache
Cassandra, MySQL). The REST approach does not set any constrains on clients or
server technologies. In this way BEANS provides one API interfaces based on REST
approach which allows to build new tools around BEANS with no changes to the
server implementation. The BEANS software has two interfaces, web and console,
but both of them are using the same REST API internally. The web and console interfaces
are clients which use JSON format to operate on the data, submit queries, view
the results and more. Almost all of the features of the web interface are
accessible also by the console interface (with some obvious limitations to
visualisation).
There are many more reasons to build software using the REST approach. It
seems that this approach fits to the BEANS software very well.

BEANS software uses only well proven technologies taken straight from the
industry. It is a very important remark. Commercial companies have been using
Big Data tools for years already. It is crucial for them to use only those tools
and software packages which are already well tested, widely used by other
entities, and have significant community of users. This increases a chance that
the companies will not loose resources in the case the software will be
abandoned by the developers. The same approaches was taken to build BEANS
software. All underlying technologies used by BEANS are already well adopted in
the industry and have a solid number of developers. BEANS uses these
technologies without changing them. It is extremely important because BEANS can
be easily updated to the newest versions whenever the underlying technologies
change. While developing BEANS a great effort was made to ensure that nothing is
developed from scratch -- instead, BEANS uses as much already available tools as
possible.

\section{Basic concepts}
\label{sec:BasicTerms}

There are only a few basic terms and definitions which users have to
familiarize themselves with in order to smoothly understand the examples
presented in Sec.~\ref{sec:Examples}. These definitions are presented here.

BEANS works as a central repository for all data the users might have.
The data are organized in \textit{datasets} and \textit{tables}. Moreover, BEANS
works also as a data analysis platform. Scripts, queries and plots for
analysis are organized in \textit{notebooks}.

A dataset is a collection of one or more tables. A table is a collection of rows
which have the same set of columns with data. Using analogy to SQL databases,
the dataset corresponds to \textit{SQL database} and the table corresponds to
\textit{SQL table}. BEANS can handle any number of datasets and every dataset
can have any number of tables. 

The dataset can be described with an arbitrary number of meta parameters. The
meta parameter is a \textit{key-value} pair. Both, the \textit{key} an
the \textit{value} are mandatory. The key can contain alphanumerical characters
only and the value might be any string of characters or a number. The meta
parameters can very much help to describe the datasets and allow to find proper datasets very
easily and instantaneously.

The datasets and tables cannot be nested. The concept of having flat list of
datasets and tables might be simple, but it is very powerful if only a proper
search engine is available. BEANS has such a search engine which allows to find
any items by name or meta parameters. The search engine finds the results
immediately, thus, there is no need to have a tree-like organization of the
datasets like folders on file systems on typical desktop computers. The meta
parameters give a much more efficient way of searching interesting items than e.g.
keeping in mind the file system locations of files.

An example one N-body simulation would be saved into one \textit{dataset}. Every
output file from the N-body simulation would be stored in a separate
\textit{table}. The dataset with the simulation would be described by the series
of \textit{meta parameters} which would correspond to the initial properties of
the N-body simulation (e.g. \textit{n = 100000} as initial number of stars,
\textit{fracb = 0.2} as initial 20\% binary fraction). In this way every dataset
with a simulation can be easily found using in a search engine queries like:
\textit{n > 40000} to find all simulations with the initial number of particles
larger than 40k, or \textit{n == 100000 AND fracb < 0.1} to find all simulations
with the initial 100k particles but with initial binary fractions smaller than 10\%. This is just
an example for a typical N-body numerical simulation. The users can arrange
their data in a any way they wish. 

A notebook is a series of items which are saved one after another and create an
integral whole. One notebook can present a complete analysis and description on
some subject. BEANS in its first public release has implemented items like
queries, texts and plots. The goal of having notebooks is to prepare a sequence
of entries in the form of a story which analyse the data. Ideally, every
notebook should be complete and self explanatory so that any other user who
opens the notebook would have everything in one place and could very easily
follow step-by-step the presented data analysis.

The first public release of BEANS contains three types of items which might be
used in notebooks. The first item allows to write queries for data analysis in
Pig Latin language (see~Sec.~\ref{sec:Technologies}). The second type of items
is text and allows to add descriptions in
Markdown\footnote{\url{https://daringfireball.net/projects/markdown/syntax}}
format (it produces HTML format from plain text). The third item type is a plot which allows
 to have basic types of diagrams like point and lines plots or
histograms. The features of the plots are very basic, their intentions is to
provide to the BEANS platform some starting mechanism to plot the data. The new
types of items, and especially the new types of plots, will be added
continuously.

The notebook can contain a list of parameters. The parameter is a key-value pair
in the same format as for datasets. But here, the parameters are used to
simplify queries and plots. For example one can define a 
parameter with a query string which finds several datasets.
Then, if one wants to make the same queries or plots but for different tables,
one can change only the parameter without changing the contents of entries in
the notebook. Before any plot or query is executed, the parameters are
substituted with the real values. It simplifies managing the notebooks. 

\section{Examples}
\label{sec:Examples}

This section presents a few examples which will show a
glimpse of the main features of the BEANS software. It is important to keep
in mind that BEANS is still under development and new features are added
continuously.
However, the first public version of BEANS contains a minimal subset of features
which already allow to help in everyday work during an analysis of
complex huge datasets.

The main purpose of the following examples is to show how one can use the BEANS
software for data analysis and plotting with easy to understand Pig Latin
language. The goal is not to analyse a real astrophysical problem.

The examples in this section work on the data produced by the MOCCA code
(see~Sec.~\ref{sec:Intro} for more details or \url{http://moccacode.net}). For
testing purposes over 300 MOCCA simulations were imported to BEANS. One MOCCA
simulation produces output consisting of several files which stores information
on the evolution of a star cluster:
global parameters like current number of stars, half-mass radius, core-radius,
number of binaries (file system.dat); data on strong dynamical interactions for
binary-binary, binary-single interactions (files inter-binbin.dat,
inter-binsin.dat); data on stellar evolution steps for binaries and single stars
(files inter-binevol.dat, inter-sinevol.dat); collisions between single stars
(file inter-coll.dat); properties of escaper objects (file escape.dat) and
snapshots in time of the entire star cluster (file snapshot.dat). Each MOCCA
simulation was saved to a separated dataset. Every file from one simulation was
imported to a separated table. Every dataset with one MOCCA simulation is
denoted with a number of meta parameters taken from the initial properties (file
mocca.ini). Among the initial properties there are the initial number of stars
(e.g. N=100000), binary fraction (e.g. fracb=0.95), initial star cluster
metallicity (e.g. zini=0.02), binary semi-major axis distribution, eccentricity
distribution, King model parameter and many more. The meta parameters allow to
quickly find any MOCCA simulation even if there would be thousands of them in
the database.
In BEANS there is no need for time consuming search for paths and files on a
local hard drive as it would have to be done with a flat files stored on a
typical PC computer.

The main window of BEANS is presented in Fig.~\ref{pic:NotebookEdit}. On the
left side there is main menu with the access to all features of BEANS: managing
datasets, notebooks, users, plugins and options. On the top bar there is a
search box to easily locate a proper notebook, dataset, or table and to view the
list of currently running jobs in the background. The main content of the web
page presented in Fig.~\ref{pic:NotebookEdit} shows edition of two entries (see
Sec.~\ref{sec:ExampleDiff} for details).

\subsection{Comparing different numerical models}
\label{sec:ExampleDiff}

The first example will provide a basic description on how to use notebooks, how to
use the edit and view modes of entries in the notebook, and how to make a simple
plot with a comparison between a
few MOCCA simulations together in one graph. 

A notebook consists of two parts. In the first part (on top) there is a menu with
an access to options and operations which may be applied to the whole notebook.
These are operations like removing or clearing the notebook, editing the title,
editing parameters of notebooks. In the notebook's menu one can also see
additional properties of the notebook (the owner, the creation date etc.) and
there are buttons to add entries to the notebook. The second part of a notebook
(below the menu) contains a series of entries.
In the first public release of BEANS there are three types of entries: text, Pig
Latin query, and plot.

The notebook can contain any number of entries of any types.
However, a good idea is to separate notebooks by its content. In this way
every notebook can be a complete analysis on some subject. Every entry has two
modes: view and edit mode. The edit mode allows to define e.g. a query or
a text and the view mode presents the results. 
The entries in edit
and view modes have their own menu with buttons which change their state and
properties. Every new entry added to the notebook is in a view state and one has
to select the entry in order to start editing it. After the edition is done, the
entry is processed. For entries like texts, the change between the view and
edit mode is fast, because changing the format from Markdown to HTML is trivial.
However, for Pig Latin queries and some plots the entry usually needs some time
to finish. When the entry is being processed, the user is informed about the
progress, and once the view is ready, the entry is automatically
reloaded.

\begin{figure*}
	\includegraphics[width=170mm]{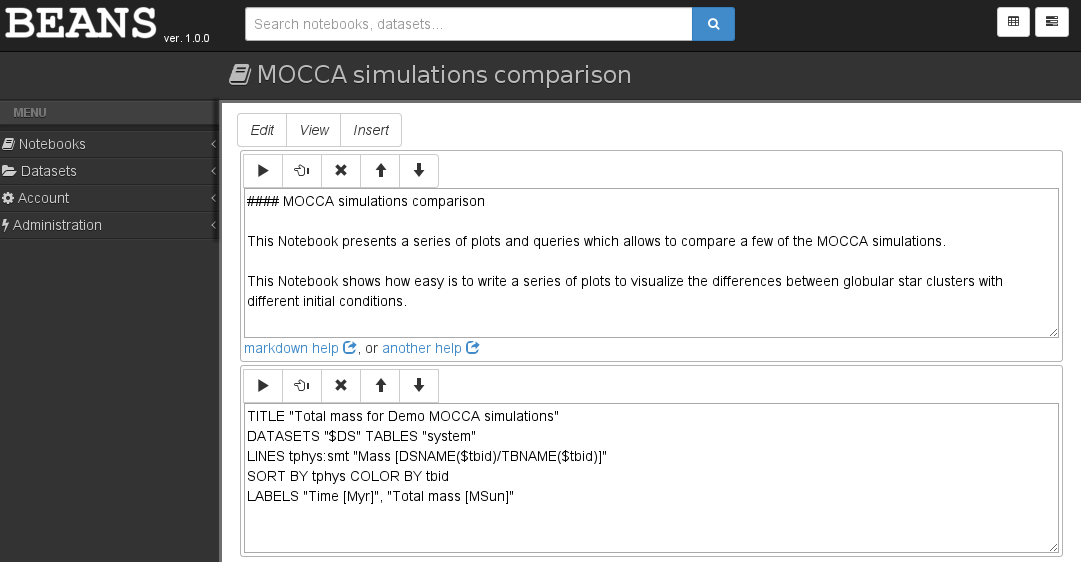}
	\caption{The figure presents BEANS main window together with an example on how to
	edit a notebook. On the left side there is the main menu with an access to all
	features of the BEANS software, on the top bar there is a search box and buttons to
	quickly find notebooks or datasets, and to examine the list and statuses of the
	currently running jobs in the background. The main content of the page
	presents a notebook in the edit mode. On the top of the notebook there is a menu with
	possible operations which can be applied to the notebook. The first editing
	entry allows to specify text in the Markdown format (see Sec.~\ref{sec:BasicTerms})
	and the second editing entry specifies a plot to compare a few tables in one graph.}
	\label{pic:NotebookEdit}
\end{figure*}

An example of a notebook in an edit mode is presented in
Fig.~\ref{pic:NotebookEdit}. The notebook has two entries:
the first entry is a text and the second entry defines a plot. The text entry is
defined using the Markdown format (see Sec.~\ref{sec:Technologies}) which as a
result produces HTML (in a view mode of the notebook, see below). The second
entry defines a plot in a simple format written especially for BEANS. Both
entries of the notebook are in the edit mode here for clarity. However, every
entry may be edited separately.

The plot in Fig.~\ref{pic:NotebookEdit} has a very simple format but it allows
to have a basic but yet powerful visualisation capabilities for the data. The
order of the statements in the plot command is arbitrary. However, the order of
statements in this example seems to be more natural and it is preferred in this
paper. Also, the new line characters do not matter for the plot, and the
statements keywords are case insensitive.
Some of the statements are optional. The full format of the plot and all
possible options are described in details in the documentation (see
Sec.~\ref{sec:License}).


The first statement in the plot example in Fig.~\ref{pic:NotebookEdit} gives the
plot a title ``Total mass for Demo MOCCA simulations'' (statement \textit{TITLE}).

The next statement in the figure specifies that the data should be read from the
datasets which can be found with the parameter \textit{``\$DS''} (statement
\textit{DATASETS}).
For this particular example the notebook parameter \$DS has a value
\textit{mocca rbar==60.0 w0==3.0 kfallb==1 n==40000 zini==0.02}. The parameter
\textit{DS} specifies the query which uses meta parameters for tables which
cover the initial conditions from the MOCCA simulations.
More precisely, the parameter specifies that the plot should find MOCCA (keyword
\textit{mocca}) simulations with tidal radius equal to 60 pc
(\textit{rbar==60.0}), King model parameter equal to 3.0 (\textit{w0==3.0}),
kick velocity during BH formation are enabled (\textit{kfallb==1}), with initial
object number of 40000 (\textit{n==40000}) and with initial solar metallicity
for star cluster (\textit{zini==0.02}).
This query finds exactly three MOCCA simulations from all of the example MOCCA
simulations imported to BEANS.

Then, the plot defines that the tables with the name
\textit{system} should be read for each of the found datasets (statement
\textit{TABLES}). The two statements together, \textit{DATASETS} and
\textit{TABLES}, find in total three \textit{system} tables from three
different datasets.

Next, there is a specification of plotting with one line (statement \textit{LINES}) of
two columns: tphys (physical time) on the X axis and smt (total star cluster mass)
on the Y axis with a label defined as \textit{``Mass
[DSNAME($tbid)/TBNAME($tbid)]''}. The label needs an additional comment on the
format. There is a number of special functions like \textit{DSNAME} and
\textit{TBNAME}.
These are the functions which get the names of datasets and tables based on the
\textit{\$tbid} parameter. The parameter \textit{\$tbid} is unique for every
table in BEANS and is added automatically to every row in a table during import.
With the special functions like \textit{DSNAME(\$tbid)}, \textit{TBNAME(\$tbid)}
one can denote the labels for the plots without specifying by hand the names of
datasets and tables. These two functions take as an argument \textit{\$tbid}
parameter and return proper names.

The statement \textit{SORT BY} sorts the rows of tables by \textit{tphys},
which is necessary to have a line plot sorted by time on X axis. 

The next statement (\textit{COLOR BY}) splits all rows read from tables based on
the \textit{tbid} column. If that statement was not used here, the plot
would have only one line with the data points taken from all three found tables.
With the statement \textit{COLOR BY} one can divide the data by any column and
plot them with separate colors very easily.

The last statement (\textit{LABELS}) in the plot command defines the labels for X
and Y axes, respectively. 

After the edition of the entries is done, one can execute the entry (with the use of the most
left button above each entry in Fig.~\ref{pic:NotebookEdit}). This will execute all
the necessary steps to convert the text into HTML (for the first entry) and to
prepare a plot (for the second entry). These tasks run in the background, and in the
meantime a user can edit some other entries in the notebook without a need
to wait.
When the view of the entries is ready, the entries will be reloaded
automatically. Depending on the amount of data to read the preparation of the
view for entries may take some time.

\begin{figure*}
	\includegraphics[width=170mm]{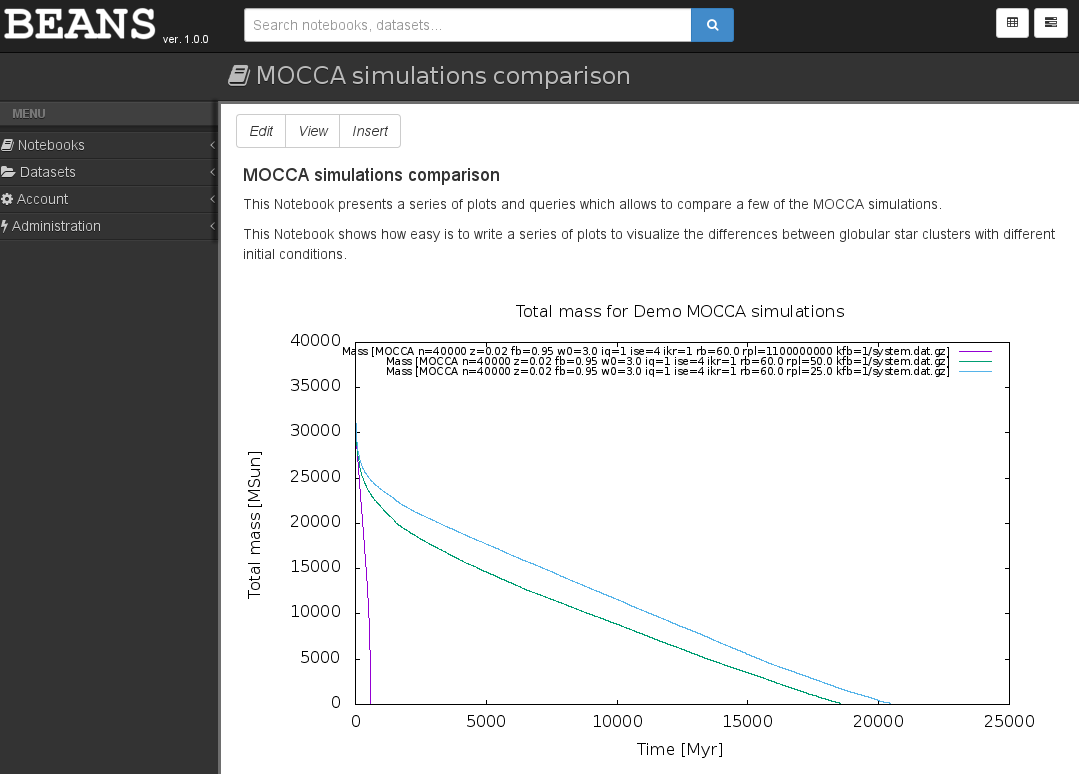}
	\caption{The figure presents a notebook in a view mode corresponding to two 
	entries edited in the Fig.~\ref{pic:NotebookEdit}. The plot shows comparison
	between three MOCCA simulations which differ with one input parameter.}
	\label{pic:NotebookView}
\end{figure*}

The notebook in the view mode, which corresponds to the text and plot entries
described above, is presented in Fig.~\ref{pic:NotebookView}. In this mode
the notebook is intended to be as distraction free as possible and thus the entries do
not show any buttons by default. The buttons with additional
viewing options are shown only after the entry is selected again. In this plot three lines are shown. Each
of them corresponds to one \textit{system} table (the labels contain the full name
of the table which is system.dat.gz) from three different datasets representing different
MOCCA simulations. Every line shows what is the overall evolution of a total
mass for a MOCCA simulation. The MOCCA simulation which evolves the fastest is the one with
rplum=1100000000. That defines an initial model for which the King model
parameter produces the standard model (i.e. not underfilled) which means that the
initial half-mass radius of the star cluster is large in comparison to the tidal
radius. This, in turn, makes the star cluster evaporates very quickly due
to the relaxation processes (these plots are not shown in this paper).

\subsection{Histograms}
\label{sec:ExampleHistograms}

The second example shows how to prepare a series of histograms with mass
distributions for three selected MOCCA simulations. Every chart will have
three histograms of masses of all stars for a particular time-step for the three
simulations. The plot will actually consist of a series of histograms -- one
histogram for one time-step. In this way it will be possible to examine how the
mass distribution changes with time for the numerical models which differ only
by one initial parameter. By putting these histograms on one plot one may
easily compare which stars are preferentially left in the star cluster
for the different models.

The basic functionalities of notebooks were discussed in the
Sec.~\ref{sec:Examples}. In the present example an additional description will be
focused on the parts of the BEANS functionalities not covered before. The
notebook for the Ex.~2 is named \textit{Mass distribution}.

\begin{figure}
	\includegraphics[width=85mm]{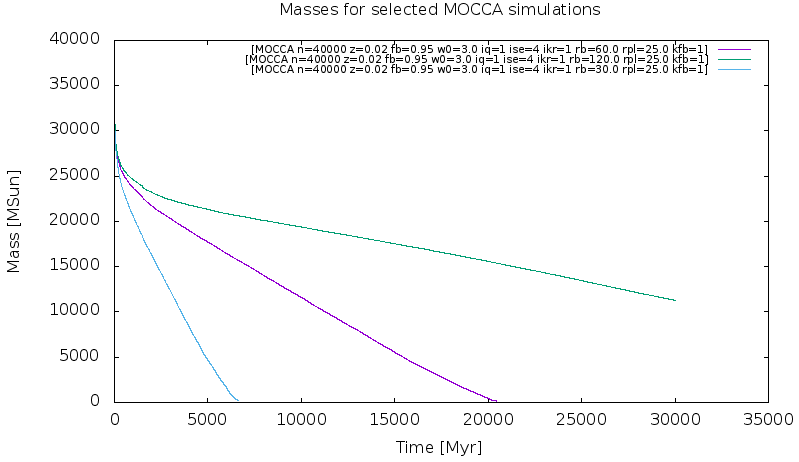}
	\caption{The figure presents the evolution of the total star cluster mass with
	time for three MOCCA simulations used in the second example showing the
	features of BEANS software. The three MOCCA simulations differ only in one
	initial parameter, rbar, which defines the tidal radius of a star cluster.
	The star cluster with the largest rbar value (tidal radius) has the slowest
	evolution because it loses less stars as a results of a relaxation processes.}
	\label{pic:Ex2Mass}
\end{figure}

In the second example also three MOCCA simulations are selected. Their total
mass evolution is presented in Fig.~\ref{pic:Ex2Mass}. The simulations differ only in one
initial parameter: rbar 
which defines the initial tidal radius. The higher the value of rbar, the
larger the initial tidal radius.
The tidal radius has a great impact on the speed of the star cluster
dissolution. Star clusters which have higher tidal radii (they are physically
further away from the host galaxy) have slower evolution. For such star
clusters less stars is ejected due to relaxation processes. Such
star cluster behaves more like an isolated cluster which does not suffer
from any external galactic potential.
The Fig.~\ref{pic:Ex2Mass} clearly shows that the star cluster with rbar==30.0
evolves the fastest (blue line) and the one with the largest rbar value evolves
the slowest (green line).

The script to produce Fig.~\ref{pic:Ex2Mass} is the following:

\begin{verbatim}
TITLE "Masses for selected MOCCA simulations"
DATASETS "$DS" TABLES "system"
LINES tphys:smt "[DSNAME($tbid)]"
SORT BY tphys
COLOR BY tbid
LABELS "Time [Myr]", "Mass [MSun]"
\end{verbatim}

The script does not use more features than in the first example. For the
explanation of the plot see Sec.~\ref{sec:ExampleDiff}. The parameter
\textit{DS} has here the value \textit{``MOCCA zini==0.02 kfallb==1 n==40000
rplum==25.0 w0==3.0''}.

To compute the histograms a Pig Latin script was created 
in the notebook. After a run of the script, 
histograms are plotted in a separate plot entry. 

\begin{figure}
	\includegraphics[width=85mm]{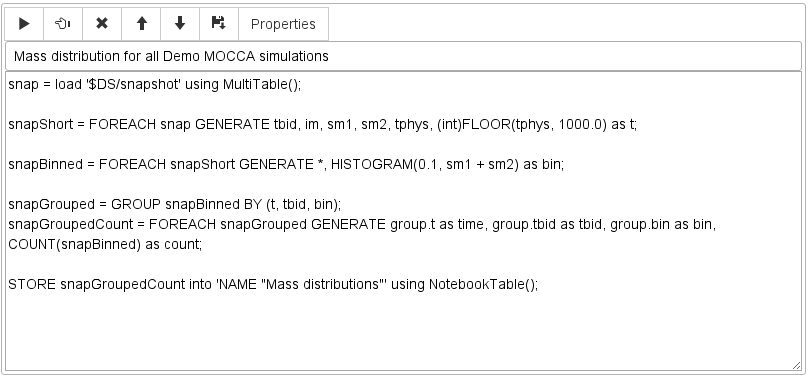}
	\caption{A notebook containing a script in Pig Latin language which produces
	values to plot histograms showing the evolution of
	mass distributions with time for three MOCCA simulations. See text for details.
	}
	\label{pic:Ex2PigHistogram}
\end{figure}

The notebook entry in edit mode with the script in Pig Latin language is
presented in Fig.~\ref{pic:Ex2PigHistogram}. It contains the script which reads the data from
three MOCCA simulations, then splits it into bins, and finally transform the data in such a
way that data are split by time-steps of the MOCCA simulations. In the end, the
data needed to plot the histogram are saved into the database. The script is
written in Pig Latin language with no changes to its format. Thus, the users
already familiar with the Apache Pig project find it clear. New
users are encouraged to read the Apache Pig documentation (see
Sec.~\ref{sec:Technologies}). For the sake of completeness the script will be
explained here step by step as a quick introduction to the Pig Latin language. 

%
%
%

The
keywords in Pig Latin language are case insensitive. However, in the examples in this paper there are capitalized for a
better clarity. 

The first statement (\textit{LOAD}) in the script
(Fig.~\ref{pic:Ex2PigHistogram}) loads the data from the BEANS database. The
data reader is called \textit{MultiTable} and was implemented in the BEANS
software to provide an access to the data seamlessly from within the Pig Latin
scripts.
The argument for the data reader is \textit{\$DS/snapshot}. It specifies the
datasets query (\textit{\$DS}) and the table query (\textit{snapshot}).
Similarly to plots, \textit{\$DS} statement, is a parameter of the notebook and
has the value \textit{``MOCCA zini==0.02 kfallb==1 n==40000 rplum==25.0
w0==3.0''}. The table query has only one keyword, \textit{snapshot}, which means
that in the found datasets the table which contains this keyword in the name
will be selected. As a result, for this very example, three MOCCA simulations
are found and in total three tables will be read from the database. All of the
rows will be read into the alias (variable) \textit{snap}. The data reader
\textit{MultiTable} allows to read rows from multiple tables, whereas the data
reader \textit{Table} would allow to read the data only from one table. If the
provided dataset and table queries
 found more than one table, the data reader \textit{Table} would raise an
exception and script would fail. In this example data are read from the tables
called \textit{snapshot}. This is a table produced by the MOCCA simulation which
contains the full snapshot in time for the entire star cluster. It contains
columns like: physical time (tphys column), ID of the stars (im column),
parameters of stars like radial position (r column), velocities (vc, vt
columns), masses (sm1, sm2 columns) etc.

The second line in the script (with statement \textit{FOREACH}) transforms all
the rows from the alias \textit{snap} by selecting only a few columns (tbid, im,
sm1, sm2, tphys) and computing two new columns (t, bin). The column
\textit{tbid}, as it was stated before, is the unique ID for every table; im is
the unique ID of a star or a binary star in MOCCA simulation; sm1, sm2 are the
masses of stars (for a single stars sm2==0); tphys is a physical time of the
simulation in Myr. The new column \textit{t} is computed with the expression
\textit{(int)FLOOR(tphys, 1000.0)}. The function FLOOR works similarly to the
floor function from C language, but here it works for any basis. It takes the
first argument (column tphys in this case) and truncates the value to the
nearest base defined with the second argument (e.g. FLOOR(100.0, 1000.0) == 0.0,
FLOOR(1100.0, 1000.0) == 1000.0, FLOOR(2900.0, 1000.0) == 2000.0 etc.). The
results is then casted to the integer value. This step is necessary to move the
different time-steps from different simulation to the same basis. Otherwise it
would not be possible to split the data with time (see additional explanation
below). The second new column, \textit{bin}, is created with a statement
\textit{HISTOGRAM(0.1, sm1 + sm2)}. The function HISTOGRAM takes two arguments.
The first one defines the width of the histogram, and the second argument
specifies the value. The function returns integer value with the bin number
where the value falls in.

The functions \textit{FLOOR} and \textit{HISTOGRAM} are examples of the
User Defined Functions (UDFs) which extends the basic Pig Latin language
with user specific functions (see Sec.~\ref{sec:Plugins} for details).

The next statement (\textit{GROUP \ldots BY}) in the script groups data by one
or more columns. In the above example the grouping is done sequentially on three
columns: t, tbid, bin. This step is needed to aggregate the values for every
time-step (t) for each MOCCA simulation separately (tbid). The result is saved
under the alias \textit{snapGrouped}.

In the next line there is again \textit{FOREACH} statement used. It computes a
few values for every group from the alias \textit{snapGrouped}. The command for
every group saves the time-step value (\textit{group.t}), ID of the simulation
(\textit{group.tbid}), bin number (\textit{group.bin}), and counts all rows which
belong to every group (\textit{COUNT}). The function \textit{COUNT} is a so-called
aggregate function. Please notice that the argument in the \textit{COUNT}
function is alias \textit{snapBinned} -- this is the alias which was used to
group the data in the previous line.
Other examples of such aggregate functions are \textit{MIN}, \textit{MAX},
\textit{SUM} (for the full list see Pig Latin docs).
The names of the columns were renamed with the statement \textit{\ldots AS
\ldots}. This part of the script is very important. It is the actual place which
computes how many rows fall to a certain bin in the histogram. It is important to
say that Apache Pig project can handle huge datasets. Thus, the same script can
build histograms for hundreds but also for billions of rows without any changes.
The only difference is the speed of calculations.

The last line in the script contains a command (\textit{STORE}) which saves the
data for histogram in a separate table for later use (e.g. plotting). The data are
saved into a table within the notebook under the name \textit{Mass
distributions}.

The script produces one table with the data for a series of histograms. The next
step is to add to the notebook a plot entry and visualise the results. The plot
script is the following:

\begin{verbatim}
TITLE "Histograms of masses"
NOTEBOOKS "Mass distribution" TABLES "mass distr"
BOXES ($bin/10.0):count "DSNAME($tbid)"
SPLIT BY time
COLOR BY tbid
LABELS "Mass [MSun]", "Number of objects"
\end{verbatim}

The majority of statements in the plot script are already explained in the Ex.~1
 (Sec.~\ref{sec:ExampleDiff}). There are only a few changes. The first one is
the statement \textit{NOTEBOOKS} which specifies that the data should be searched in
the notebook which contains the keywords \textit{Mass distribution} (in the
Ex.~1 there was a statement \textit{DATASETS}). 

The second change is the type of the plot -- \textit{BOXES}. This types of a
plot produces histograms, in this example, from the columns \textit{bin} (X
axis) and \textit{count} (Y axis). The X axis is actually defined as a
mathematical expression which uses the column \textit{bin}. In mathematical
expressions the column names have to be prefixed with a dollar sign to
mark them as variables and the whole expression has to be surrounded with the
parentheses.

The last new statement is \textit{SPLIT BY}. This is the command which produces
a series of plots splitting all the data by the column \textit{time}. For every
value of the column \textit{time} there will be made one separate plot.

\begin{figure*}
	\includegraphics[width=160mm]{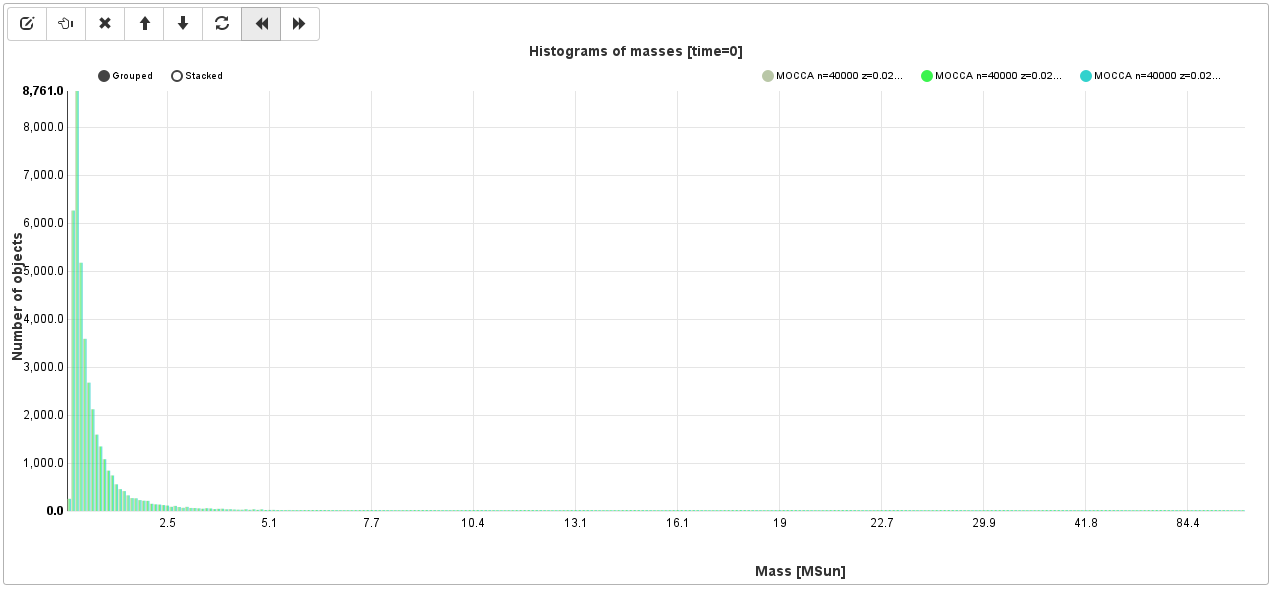}
	\includegraphics[width=160mm]{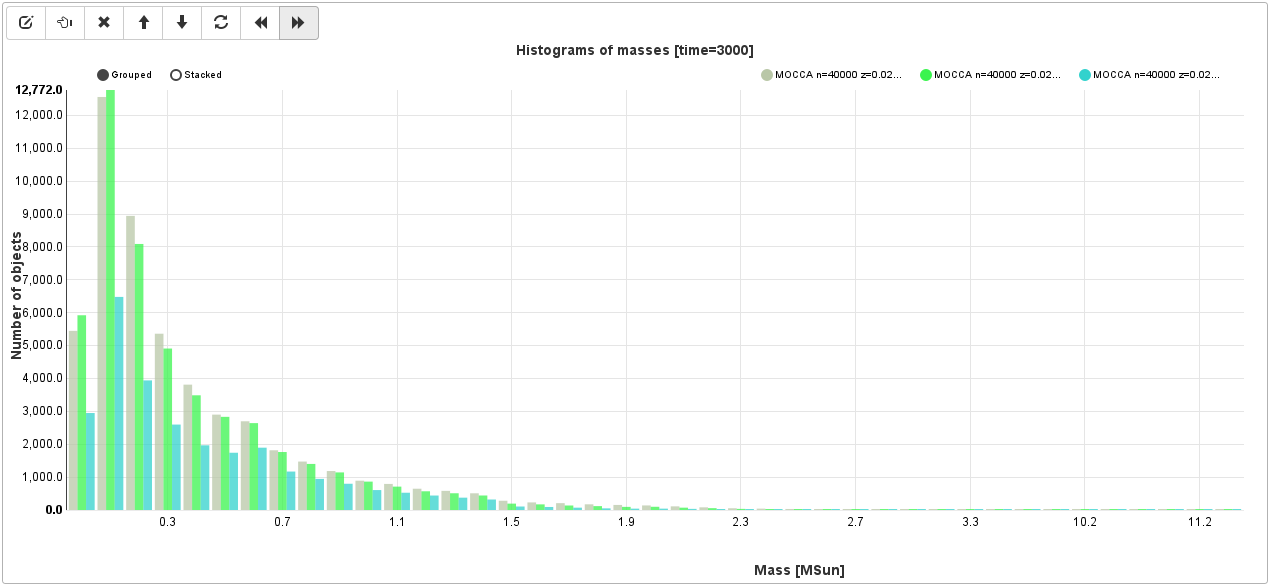}
	\includegraphics[width=160mm]{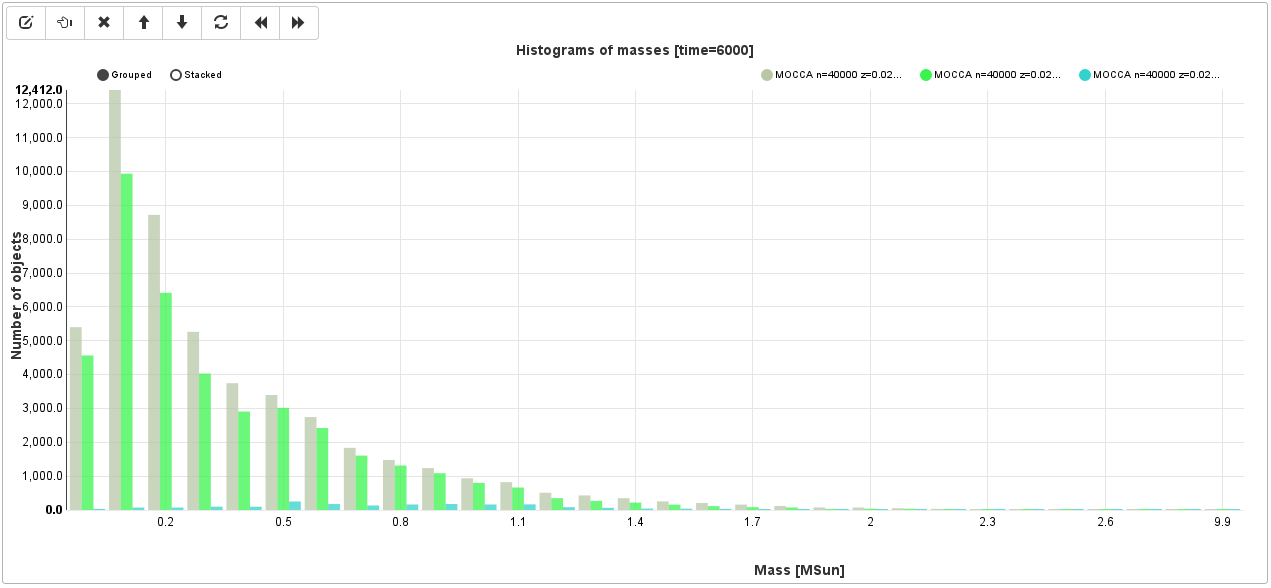}
	\caption{Histograms with the mass distribution for three selected MOCCA
	simulations for three time-steps of 0~Myr, 3000~Myr and 6000~Myr. The simulations
	have the same initial conditions except for the different tidal radius (see text for
	details). The three plots are contained within one plot entry in a notebook.
	The plots are switched using the left- and right-arrows from the menu. The
	MOCCA simulations have exactly the same initial conditions, thus the plot for
	time 0 is the same for all three.}
	\label{pic:Ex2PlotHistogram}
\end{figure*} 

The resulting histograms for the script are shown in the
Fig.~\ref{pic:Ex2PlotHistogram}. Every plot contains the mass
distributions for three selected MOCCA simulations which differ in one
parameter (tidal radius). The figure shows from the top the
histograms for time-steps of 0~Myr, 3000~Myr and 6000~Myr. The first histogram is
identical for three simulations because the MOCCA code started with exactly the
same initial mass functions. After already 3~Gyr and clearly after 6~Gyr one can
see that for the lower tidal radii (rbar==60.0 and rbar=30.0) there is less low
massive stars. These are the stars preferentially ejected from star clusters
due to relaxation processes or strong dynamical interactions. In turn, for the
simulation with rbar==120.0 (the largest tidal radius of the three
simulations), the overall evolution of the star cluster is much slower, and there
is still a significant number of low massive stars after 6~Gyr. 

Plots for the next time-steps can be viewed with the left- and right-arrows
visible on the right side in the menu. In this work we present only three
histograms, but there are next plots available up to 30~Gyr (see
Fig.~\ref{pic:Ex2Mass}).

\subsection{Multi-parameter plot}
\label{sec:ExampleParallel}

The last example (Ex.~3) will show the capabilities of powerful plots made with
the library D3 (see Sec.~\ref{sec:Technologies}). In the first public
release BEANS is integrated only with the plot named
\textit{Parallel coordinates}. It allows to interactively examine relations
between many parameters on a single plot. The D3 plots are non-standard
but extremely powerful, clean and easy to understand. 

In the Ex.~3 we present a script and the resulting plot which shows
interactively how the basic global properties from all MOCCA simulations
with n==100000 change with time every 400~Myr. On a single plot it is
possible to see e.g.
how the actual number of binaries in the star cluster relates to half-mass
radii, number of stars or to a few other parameters.  

First, one has to prepare a script which reads all of the MOCCA simulations
which start with initial n==100000 and then groups them by the same time-steps.
The script is the following: 

\begin{verbatim}
%DECLARE step '400.0';

rows = LOAD 'MOCCA n==100000/system' using MultiTable();
rows = FILTER rows BY tphys < 14000;

rows = FOREACH rows GENERATE tbid, nt, DSPARAM(DSID(tbid), 
'zini') as zini, DSPARAM(DSID(tbid), 'fracb') as fracb, 
rchut2, r_h as rh, rtid, xrc, vc, u1, irc, tphys, 
FLOOR(tphys, $step) as t, tphys - FLOOR(tphys, $step) 
as diff;
   
rowsGrouped = GROUP rows BY (t, tbid);

rowsFlat = FOREACH rowsGrouped GENERATE group.tbid as 
tbid, group.t as t, FLATTEN(rows), MIN(rows.diff) as 
minDiff; 

rowsFlat = FOREACH rowsFlat GENERATE tbid, (int)t, 
rows::tphys as tphys, rows::diff as diff, minDiff, 
rows::nt as nt, rows::zini as zini, rows::fracb as
fracb, rows::rchut2 as rchut2, rows::rh as rh, 
rows::rtid as rtid, rows::xrc as xrc, rows::vc as vc, 
rows::u1 as u1, rows::irc as irc;

rowsFlat = FILTER rowsFlat BY tphys - t < minDiff + 0.001;

STORE rowsFlat into 'NAME "System every 400M"' using 
NotebookTable();
\end{verbatim}

The script is in many places similar to the script in Ex.~2 (see
Sec.~\ref{sec:ExampleHistograms}). A new statement is in the first line --
\textit{DECLARE}.
It defines a parameter inside a Pig Latin script.
It is used in the script in two places.
The script then reads tables \textit{system} from all MOCCA simulations with
initial number of stars equal to 100000. In the next line the rows are filtered
and only those for which physical time is less than 14~Gyr (\textit{tphys <
14000}) are left.

The next line with the \textit{FOREACH} statement reads only several columns from
every row and adds a few more columns. The columns which are only rewritten are
e.g.
rtid, xrc, vc, u1. The column \textit{zini} is computed with the expression
\textit{DSPARAM(DSID(tbid), 'zini')}. This expression reads the value of
\textit{zini} from the meta parameter defined for the MOCCA
simulation during import (see Sec.~\ref{sec:BasicTerms} for detailed description
of meta parameters).
The functions \textit{DSID} and \textit{DSPARAM} are UDFs -- helper functions
written for BEANS (see Sec.~\ref{sec:Plugins} for details on writing own
functions). The function \textit{DSID} takes one argument, ID of a table
\textit{tbid}, and returns the ID of a dataset to which the table belongs.
Finally, the function \textit{DSPARAM} takes two arguments (dataset ID and
a name of a meta parameter) and returns the value of the meta parameter. In this way one can
get the values from meta parameter describing the
datasets.

In the same line two additional columns, \textit{t} and \textit{diff}, are
calculated using the \textit{FLOOR} function already introduced in the Ex.~2.
Different MOCCA simulations have different time-steps. Thus, the lines in the
output files are denoted with different physical times. One simulation can have
the time 402.01~Myr, whereas another 404.99~Myr. They have to be moved to
the same basis for grouping.
These columns will be used to combine MOCCA simulations by the physical times
which are the closes to each other.

The next two lines in the script group the rows by time (\textit{t}) and table
ID (\textit{tbid}) and then aggregate the groups with the \textit{MIN} function.
The statement \textit{FLATTEN} is used to copy all columns from alias
\textit{rows} to every permutation of groups \textit{group.tbid} and
\textit{group.tbid}. Without that statement the number of rows in the alias
\textit{rowsFlat} would be equal to the number of groups created by the previous
statement. With the command \textit{FLATTEN} the number of rows will not
change, but additionally every row will get the \textit{group.tbid},
\textit{group.tbid}, and a value from the aggregate function \textit{MIN}.

The next line only
renames the columns. The following line actually leaves only those rows
which are closest to the base time (\textit{t}) -- other time-steps are removed
from the results. In this way one can have values like core radius (rchut2),
central velocity (vc) for every 400.0~Myr no matter what actual time-steps
for different MOCCA simulations are. The last line of the script saves the
results into the table \textit{System every 400M}.

\begin{figure*}
	\includegraphics[width=170mm]{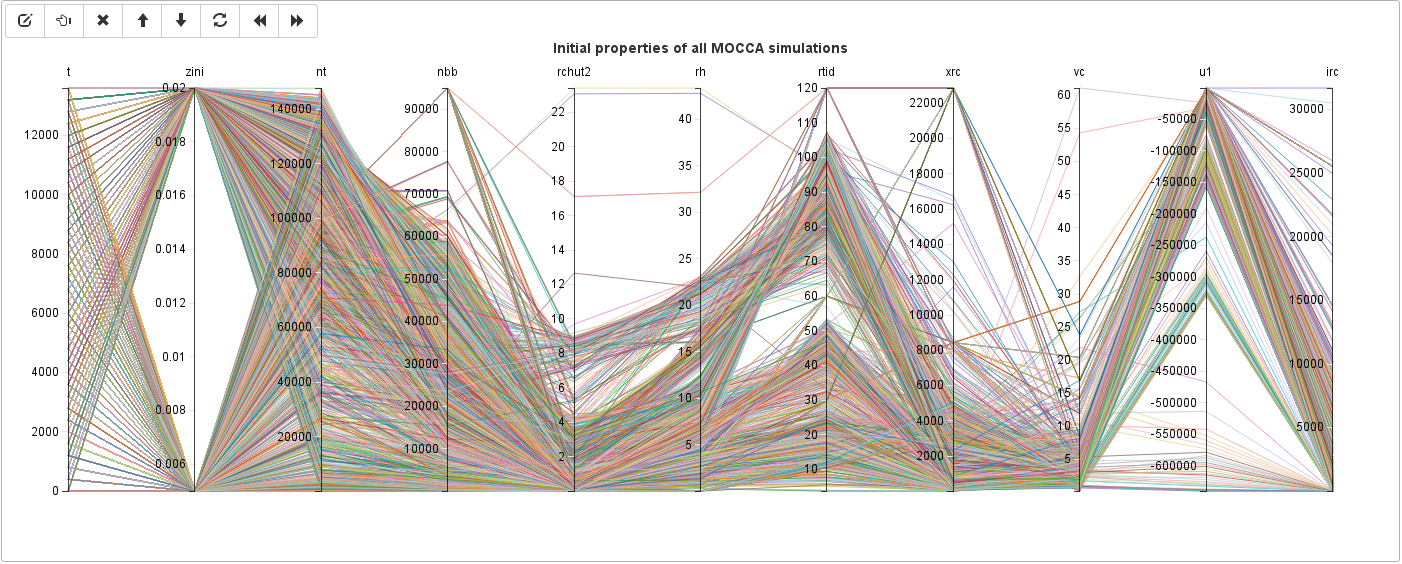}
	\includegraphics[width=170mm]{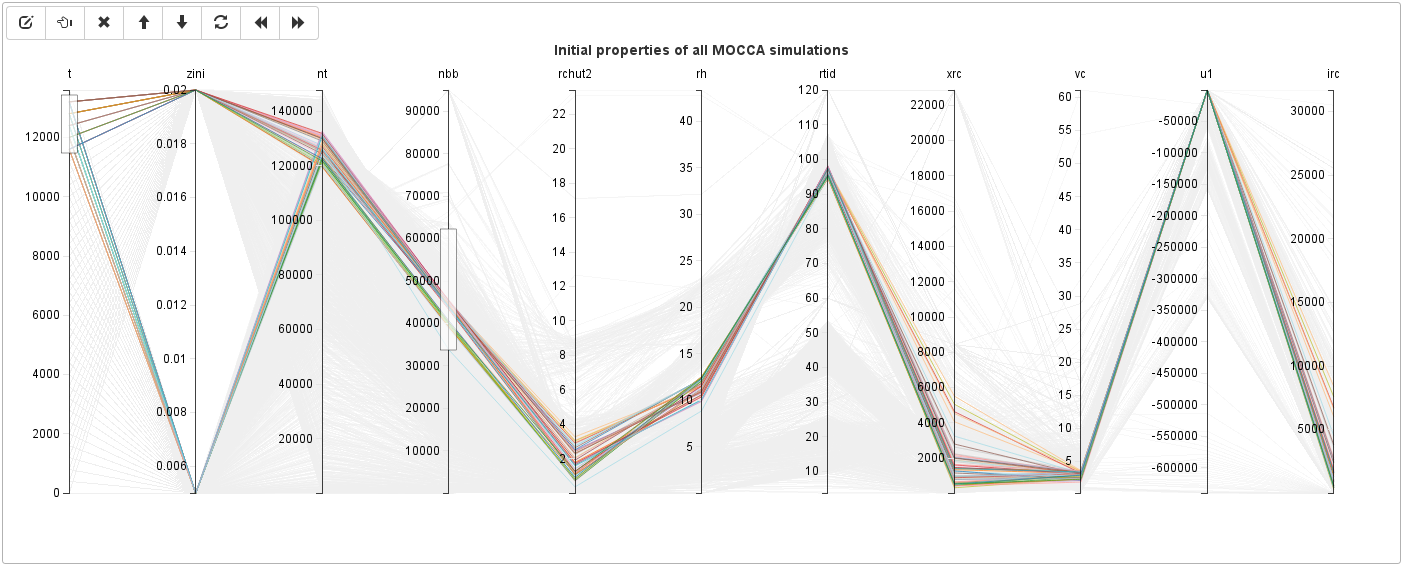}
	\caption{Example of an interactive D3 plot which presents how a set of global
	parameters of many MOCCA simulations relates to each other for every 400~Myr.
	On the top panel all lines are plotted, one line per one MOCCA simulation per
	one time-step. On the bottom panel only the subset of columns t and nbb is
	selected to show that one can constrain the values interactivelly. The
	parameters are the following: physical time in Myr (t), initial metallicity (zini), current number of
stars (nt), binaries (nbb), core radius (rchut2), half-mass radius (rh), tidal
radius (rtid), core mass (xrc), central velocity (vc), central potential (u1)
and number of stars in the core (irc).}
	\label{pic:Ex3Plot}
\end{figure*} 

The D3 plot which presents the results of the script is shown in 
Fig.~\ref{pic:Ex3Plot}. It contains a number of parameters describing properties
of star clusters like initial metallicity (zini), current number of stars (nt),
binaries (nbb), core radius (rchut2), half-mass radius (rh), tidal radius
(rtid), core mass (xrc), central velocity (vc), central potential (u1) and
number of stars in the core (irc). On the top panel all lines for all MOCCA
simulations are shown.
The plots are interactive and on the bottom panel only a subset of values for
columns t (around 13~Gyr) and nbb (around 50000) is selected (transparent
rectangles on the vertical axes).
Each line corresponds to one MOCCA simulation for a particular time-step. By
moving the selection rectangles one can see interactively which lines are
becoming visible. On one plot one can compare many parameters and  examine
how changes in one parameter influence the others. In the bottom panel there are
selected star clusters around 13~Gyr old and which have still around 50\% of
binaries. The plot shows that these parameters are characteristic only for star
cluster which have large tidal radii (rtid $\sim$ 90.0) and thus are almost
isolated star clusters, and have large half-mass radii (rh $\sim$ 10) and
thus have low concentrations.

The plot was made with the following script:

\begin{verbatim}
TITLE "Initial properties of all MOCCA simulations"
NOTEBOOKS "Parallel coordinates" TABLES "System every 1 Gyr"
PARALLEL t:zini:nt:nbb:rchut2:rh:rtid:xrc:vc:u1:irc
LABELS "Parameters", "Values of the parameters"
\end{verbatim}

The only new statement in the plot is \textit{PARALLEL} which defines a series
of columns separated with colons.   

Fig.~\ref{pic:Ex3Plot} shows the power of D3 visualization library. One of
the main challenges in data visualization is an efficient and clear
representation of multidimensional data. Human perception really struggles to read data from plots
with more than 3 dimensions. The figure demonstrates a plot called
\textit{Parallel coordinates} which can be used to visualise the relations
between a number of parameters, interactively on one plot. This is just one
example of a D3 plot. There is already a
large number of very different D3 plots created by the
community available
for download\footnote{\url{https://github.com/mbostock/d3/wiki/Gallery}}. BEANS
will be gradually integrated with more and more such powerful plots.

Notebook can contain any number of text, plot and Pig Latin
entries. By creating a series of entries one can
create a story which from scratch explains all steps in data analysis on a
particular subject. What is more, one can create a generic notebook which contains a series
of entries for easy comparison of a selected set of tables (e.g. defined by the
notebook's parameter).

\section{Plugins}
\label{sec:Plugins}

BEANS allows to extend its basic set of features with plugins. In the first
public version plugins allow to extend only Pig Latin language with User
Defined Functions (UDFs). A few examples of UDFs were used already in
Sec.~\ref{sec:ExampleHistograms} and \ref{sec:ExampleParallel}. UDFs allow to
extends BEANS to the users' specific needs. Moreover, they increase
the clarity of the scripts by replacing repeatable, complex data manipulations with
simpler to use functions.

Currently, there are two plugins written for BEANS. The first one is written for
simplifying the analysis of the MOCCA simulations. It contains a series of
functions which help to extract and transform some columns from the very complex
output of the MOCCA simulations. It contains also a few commands which create
notebooks with preconfigured set of plots (around 30) and queries which
helps and speeds up significantly the comparison of different models. One can
create a notebook, change a parameter and then only reload all
plots to see the actual differences between some models.

The second plugin is a Gaia Archive Core Systems (GACS) plugin. Its purpose is
to provide a connection to the GACS web
portal\footnote{\url{http://gaia.esac.esa.int/archive/}}. This portal is a
starting point to access the archive of the Gaia mission. The plugin allows to
write ADQL query which calls internally the GACS portal. After the query is
finished, the data are automatically imported to BEANS for analysis. The goal of
this plugin is to simplify comparison of complex numerical models with the Gaia
archive. This plugin will be described in more details in the next paper where
BEANS will be used to compare a numerical model of the Galaxy with the entire
Gaia catalogue (currently a numerical model of the Galaxy filled with
objects which Gaia will potentially observe).

Plugins are quite easy to write for everyone with moderate programming skills.
UDFs for Pig Latin language one can write currently in Java, Python/Jython,
JavaScript, Ruby and Groovy languages. There is also available a public
repository with complex and very useful scripts for Apache Pig project called
Piggy~Bank\footnote{\url{https://cwiki.apache.org/confluence/display/PIG/PiggyBank}}.
For the technical description about writing plugins see links in
Sec.~\ref{sec:License}.

\section{Future plans}
\label{sec:FuturePlans}

The BEANS software is under development and more features are being added
continuously. BEANS is planned to be developed taking seriously into
consideration a feedback from the community. Before it happens, a few new
features planned for the next releases are described below.

Currently BEANS allows to have multiple users. However, every user has access to
any data stored in the BEANS database. Simple user rights management is planned
for implementation as soon as possible in order to be able to create e.g. private
notebooks or datasets.

BEANS allows to analyse only the data which are imported to BEANS
software (into Apache Cassandra database). There are planned several additional
connectors allowing Pig Latin scripts to read the files straight from the local
hard drives. In this way it will not be mandatory to import the data into BEANS
for analysis. For some users it may be not feasible to import all the data into
BEANS. Moreover, we plan to expand supported input files format,
especially to csv, HDF5 and FITS files.

BEANS tracks internally the list of tables used as an input for all the
plots and Pig Latin queries. In the case of uploading more datasets into BEANS
it may happen that the same query or a plot could read the data from more
tables. We plan to add a feature to track which queries and plots need an update.
The user would be asked for confirmation and such entries would be reloaded
automatically. Similarly, in the case of removal of some tables, the
user would be notified about an obsolete entry and asked for an update. This
mechanism would ensure to keep all the notebooks up-to-date as if they would be
``living notebooks''.

BEANS will have more types of entries in notebooks. In the present version there
are present three types only: text, plot, and Pig script. New types will be added
continuously but among the most important there are entries able to run
python (e.g. matplotlib\footnote{\url{http://matplotlib.org/}} library), AWK and
gnuplot\footnote{\url{http://www.gnuplot.info/}} codes. More types will be added
depending on which language or tool will be most needed by the community.
Additionally, BEANS will be integrated with more types of D3 plots.

The scalability of BEANS will be tested in the next paper where the Gaia archive
will be compared with a complex numerical simulation of the Galaxy (TBs of
data).
It will also be the publication introducing in more detail the GACS plugin
connecting to the Gaia archive.
BEANS will be tested on a cluster of Apache Cassandra nodes and Apache Hadoop
nodes, whereas examples presented in the current publication were executed on a
computer with one instance of Apache Cassandra and Elastic. The Hadoop jobs were
working in embedded mode (only on one machine). The purpose of this work was to
present BEANS software and to show how it helps to handle many different
datasets and tables from within one piece of software. In our following work we
will present how BEANS is performing for really huge datasets.

The BEANS software is not written
for any automatic data analysis. It has not got implemented any machine learning
techniques. It will not find correlations in the data
automatically. BEANS is also not a good solution for a user looking for instant,
almost real-time analytics. The best use of the BEANS software
is a supervised, incremental data analysis -- at least in the nearest future.



\section{License and links}
\label{sec:License}

BEANS software uses in its core only Open Source libraries and technologies.
The code of the BEANS itself will be published under the Open Source license too
-- Apache License, Version 2.0\footnote{\url{http://www.apache.org/licenses/}},
January 2004. Apache License 2.0 is very liberal license and allows to use software freely.

The description of the BEANS software, features, source code and installation
procedure one can find on the web page \url{http://beanscode.net}. More
technical description, list of bugs, new features requests, instructions on how to
write own plugins, and many more, one can find on the page
\url{http://redmine.beanscode.net}.


\section{Summary}
\label{sec:Summary}

The BEANS software is a complete attempt to incorporate into
science, in particular into astronomy, data analysis techniques which already
became mainstream in the industry. Predictions for data analysis suggest that
solution based on Apache Hadoop will be dominating. BEANS software (or similar
solutions) based on these technologies can truly simplify everyday work of
scientists.

Most prominent features of the BEANS software are an easy data management,
powerful and clean Pig Latin language for querying, filtering the data, and an
easy to use plotting library. These allow to have one web/cli interface to the
data, no matter how many datasets and tables are present in the database. What
is more, it is reasonably easy for users with a moderate technical knowledge to
increase an overall size or computing power of a cluster. To expand such cluster
one has to add new nodes, and the cluster will automatically redistribute the
data evenly to all nodes.
The installation is also relatively easy. The only significant requirement for
the users is to learn the Pig Latin language for data analysis. This language
differs from popular ones like python, C, or Java, but on the other hand the
documentation with examples of Pig Latin language is short. If a user invests a
time to learn the Pig Latin, BEANS will provide the complete solution for
storage and analysis of data of any size.

The data analysis is organized in notebooks which consists of a sequence of
text, queries and plots. Thus, every notebook may contain an integral and complete
study on some phenomena. The notebooks, datasets and tables are easy to find
with powerful search engine which means that even for thousands of objects there
is no problem in finding any piece of information in the entire database. 


\section*{Acknowledgment}

The research leading to these results has received funding
from the European Community's Seventh Framework Programme (FP7-SPACE-2013-1)
under grant agreement n.~606740.

The project was supported partially by Polish National Science Center grants
DEC-2011/01/N/ST9/06000 and DEC-2012/07/B/ST9/04412.





\bibliographystyle{mnras} 
\bibliography{beans}

\begin{thebibliography}{}
\makeatletter
\relax
\def\mn@urlcharsother{\let\do\@makeother \do\$\do\&\do\#\do\^\do\_\do\%\do\~}
\def\mn@doi{\begingroup\mn@urlcharsother \@ifnextchar [ {\mn@doi@}
  {\mn@doi@[]}}
\def\mn@doi@[#1]#2{\def\@tempa{#1}\ifx\@tempa\@empty \href
  {http://dx.doi.org/#2} {doi:#2}\else \href {http://dx.doi.org/#2} {#1}\fi
  \endgroup}
\def\mn@eprint#1#2{\mn@eprint@#1:#2::\@nil}
\def\mn@eprint@arXiv#1{\href {http://arxiv.org/abs/#1} {{\tt arXiv:#1}}}
\def\mn@eprint@dblp#1{\href {http://dblp.uni-trier.de/rec/bibtex/#1.xml}
  {dblp:#1}}
\def\mn@eprint@#1:#2:#3:#4\@nil{\def\@tempa {#1}\def\@tempb {#2}\def\@tempc
  {#3}\ifx \@tempc \@empty \let \@tempc \@tempb \let \@tempb \@tempa \fi \ifx
  \@tempb \@empty \def\@tempb {arXiv}\fi \@ifundefined
  {mn@eprint@\@tempb}{\@tempb:\@tempc}{\expandafter \expandafter \csname
  mn@eprint@\@tempb\endcsname \expandafter{\@tempc}}}

\bibitem[\protect\citeauthoryear{{Dean} \& {Ghemawat}}{{Dean} \&
  {Ghemawat}}{2008}]{Dean1327452.1327492}
{Dean} J.,  {Ghemawat} S.,  2008, \mn@doi [Commun. ACM]
  {10.1145/1327452.1327492}, 51, 107

\bibitem[\protect\citeauthoryear{{Giersz}, {Heggie}, {Hurley}  \&
  {Hypki}}{{Giersz} et~al.}{2013}]{Giersz2013MNRAS.431.2184G}
{Giersz} M.,  {Heggie} D.~C.,  {Hurley} J.~R.,   {Hypki} A.,  2013, \mn@doi
  [\mnras] {10.1093/mnras/stt307}, \href
  {http://adsabs.harvard.edu/abs/2013MNRAS.431.2184G} {431, 2184}

\bibitem[\protect\citeauthoryear{{Giersz}, {Leigh}, {Hypki}, {L{\"u}tzgendorf}
  \& {Askar}}{{Giersz} et~al.}{2015}]{Giersz2015MNRAS.454.3150G}
{Giersz} M.,  {Leigh} N.,  {Hypki} A.,  {L{\"u}tzgendorf} N.,   {Askar} A.,
  2015, \mn@doi [\mnras] {10.1093/mnras/stv2162}, \href
  {http://adsabs.harvard.edu/abs/2015MNRAS.454.3150G} {454, 3150}

\bibitem[\protect\citeauthoryear{{Hypki} \& {Giersz}}{{Hypki} \&
  {Giersz}}{2013}]{Hypki2013MNRAS.429.1221H}
{Hypki} A.,  {Giersz} M.,  2013, \mn@doi [\mnras] {10.1093/mnras/sts415}, \href
  {http://adsabs.harvard.edu/abs/2013MNRAS.429.1221H} {429, 1221}

\bibitem[\protect\citeauthoryear{{Ivezic} et~al.,}{{Ivezic}
  et~al.}{2008}]{2008arXiv0805.2366I}
{Ivezic} Z.,  et~al., 2008, preprint, \href
  {http://adsabs.harvard.edu/abs/2008arXiv0805.2366I} {} (\mn@eprint {arXiv}
  {0805.2366})

\bibitem[\protect\citeauthoryear{{Lindegren} et~al.,}{{Lindegren}
  et~al.}{2008}]{Gaia2008IAUS..248..217L}
{Lindegren} L.,  et~al., 2008, in {Jin} W.~J.,  {Platais} I.,   {Perryman}
  M.~A.~C.,  eds,  IAU Symposium Vol. 248, IAU Symposium. pp 217--223,
  \mn@doi{10.1017/S1743921308019133}

\makeatother
\end{thebibliography}



\appendix

%
%
%





\bsp	
\label{lastpage}
\end{document}